\documentclass[%
reprint,
nofootinbib,
amsmath,amssymb,
aps,
prl,
floatfix,
superscriptaddress
]{revtex4-2}

\usepackage{graphicx}
\usepackage{dcolumn}
\usepackage{bm}
\usepackage{hyperref}
\usepackage{comment}
\usepackage{xcolor}
\usepackage[caption=false]{subfig}
\usepackage{siunitx}
\usepackage{multirow}
\newcommand\T{\rule{0pt}{\dimexpr\baselineskip+12pt\relax}} 
\newcommand\B{\rule[-\dimexpr0.3\baselineskip+6pt\relax]{0pt}{0pt}}

\begin{document}


\preprint{APS/123-QED}

\title{NOvA Dual-Baseline Search for Active-to-Sterile Neutrino Oscillations using Neutrino- and Antineutrino-Enriched Samples}

\newcommand{\ANL}{Argonne National Laboratory, Argonne, Illinois 60439, 
USA}
\newcommand{\Bandirma}{Bandirma Onyedi Eyl\"ul University, Faculty of 
Engineering and Natural Sciences, Engineering Sciences Department, 
10200, Bandirma, Balıkesir, Turkey}
\newcommand{\ICS}{Institute of Computer Science, The Czech 
Academy of Sciences, 
182 07 Prague, Czech Republic}
\newcommand{\IOP}{Institute of Physics, The Czech 
Academy of Sciences, 
182 21 Prague, Czech Republic}
\newcommand{\Atlantico}{Universidad del Atlantico,
Carrera 30 No.\ 8-49, Puerto Colombia, Atlantico, Colombia}
\newcommand{\BHU}{Department of Physics, Institute of Science, Banaras 
Hindu University, Varanasi, 221 005, India}
\newcommand{\UCLA}{Physics and Astronomy Department, UCLA, Box 951547, Los 
Angeles, California 90095-1547, USA}
\newcommand{\Caltech}{California Institute of 
Technology, Pasadena, California 91125, USA}
\newcommand{\CUSB}{Central University of South Bihar, 
NH-120, Gaya Panchanpur Road, Post Fatehpur, 
Gaya, Bihar, 824 236, India}
\newcommand{\Charles}
{Charles University, Faculty of Mathematics and Physics,
 Institute of Particle and Nuclear Physics, Prague, Czech Republic}
\newcommand{\Cincinnati}{Department of Physics, University of Cincinnati, 
Cincinnati, Ohio 45221, USA}
\newcommand{\CSU}{Department of Physics, Colorado 
State University, Fort Collins, CO 80523-1875, USA}
\newcommand{\CTU}{Czech Technical University in Prague,
Brehova 7, 115 19 Prague 1, Czech Republic}
\newcommand{\Dallas}{Physics Department, University of Texas at Dallas,
800 W. Campbell Rd. Richardson, Texas 75083-0688, USA}
\newcommand{\DallasU}{University of Dallas, 1845 E 
Northgate Drive, Irving, Texas 75062 USA}
\newcommand{\Delhi}{Department of Physics and Astrophysics, University of 
Delhi, Delhi 110007, India}
\newcommand{\JINR}{Joint Institute for Nuclear Research,  
Dubna, Moscow region 141980, Russia}
\newcommand{\Erciyes}{
Department of Physics, Erciyes University, Kayseri 38030, Turkey}
\newcommand{\FNAL}{Fermi National Accelerator Laboratory, Batavia, 
Illinois 60510, USA}
\newcommand{\FSU}{Florida State University, Tallahassee, Florida 32306, USA}
\newcommand{\UFG}{Instituto de F\'{i}sica, Universidade Federal de 
Goi\'{a}s, Goi\^{a}nia, Goi\'{a}s, 74690-900, Brazil}
\newcommand{\Guwahati}{Department of Physics, IIT Guwahati, Guwahati, 781 
039, India}
\newcommand{\Harvard}{Department of Physics, Harvard University, 
Cambridge, Massachusetts 02138, USA}
\newcommand{\Homi}{Homi Bhabha National Institute, Training School Complex,
 Anushakti Nagar, Mumbai 400094, India}
\newcommand{\Houston}{Department of Physics, 
University of Houston, Houston, Texas 77204, USA}
\newcommand{\IHyderabad}{Department of Physics, IIT Hyderabad, Hyderabad, 
502 205, India}
\newcommand{\Hyderabad}{School of Physics, University of Hyderabad, 
Hyderabad, 500 046, India}
\newcommand{\IIT}{Illinois Institute of Technology,
Chicago IL 60616, USA}
\newcommand{\Imperial}{Imperial College London, Department of Physics,
 London, United Kingdom}
\newcommand{\Indiana}{Indiana University, Bloomington, Indiana 47405, 
USA}
\newcommand{\INR}{Institute for Nuclear Research of Russia, Academy of 
Sciences 7a, 60th October Anniversary prospect, Moscow 117312, Russia}
\newcommand{\UIowa}{Department of Physics and Astronomy, University of Iowa, 
Iowa City, Iowa 52242, USA}
\newcommand{\ISU}{Department of Physics and Astronomy, Iowa State 
University, Ames, Iowa 50011, USA}
\newcommand{\Irvine}{Department of Physics and Astronomy, 
University of California at Irvine, Irvine, California 92697, USA}
\newcommand{\Jammu}{Department of Physics and Electronics, University of 
Jammu, Jammu Tawi, 180 006, Jammu and Kashmir, India}
\newcommand{\Lebedev}{Nuclear Physics and Astrophysics Division, Lebedev 
Physical 
Institute, Leninsky Prospect 53, 119991 Moscow, Russia}
\newcommand{\Magdalena}{Universidad del Magdalena, Carrera 32 No 22-08 Santa Marta, Colombia}
\newcommand{\MSU}{Department of Physics and Astronomy, Michigan State 
University, East Lansing, Michigan 48824, USA}
\newcommand{\Crookston}{Math, Science and Technology Department, University 
of Minnesota Crookston, Crookston, Minnesota 56716, USA}
\newcommand{\Duluth}{Department of Physics and Astronomy, 
University of Minnesota Duluth, Duluth, Minnesota 55812, USA}
\newcommand{\Minnesota}{School of Physics and Astronomy, University of 
Minnesota Twin Cities, Minneapolis, Minnesota 55455, USA}
\newcommand{\Mississippi}{University of Mississippi, University, Mississippi 38677, USA}
\newcommand{\NISER}{National Institute of Science Education and Research, 
Bhubaneswar, Khurda, Odisha 752050, India}
\newcommand{\OSU}{Department of Physics, Ohio State University, Columbus,
Ohio 43210, USA}
\newcommand{\Oxford}{Subdepartment of Particle Physics, 
University of Oxford, Oxford OX1 3RH, United Kingdom}
\newcommand{\Panjab}{Department of Physics, Panjab University, 
Chandigarh, 160 014, India}
\newcommand{\Pitt}{Department of Physics, 
University of Pittsburgh, Pittsburgh, Pennsylvania 15260, USA}
\newcommand{\QMU}{Particle Physics Research Centre, 
Department of Physics and Astronomy,
Queen Mary University of London,
London E1 4NS, United Kingdom}
\newcommand{\RAL}{Rutherford Appleton Laboratory, Science 
and 
Technology Facilities Council, Didcot, OX11 0QX, United Kingdom}
\newcommand{\SAlabama}{Department of Physics, University of 
South Alabama, Mobile, Alabama 36688, USA} 
\newcommand{\Carolina}{Department of Physics and Astronomy, University of 
South Carolina, Columbia, South Carolina 29208, USA}
\newcommand{\SDakota}{South Dakota School of Mines and Technology, Rapid 
City, South Dakota 57701, USA}
\newcommand{\SMU}{Department of Physics, Southern Methodist University, 
Dallas, Texas 75275, USA}
\newcommand{\Stanford}{Department of Physics, Stanford University, 
Stanford, California 94305, USA}
\newcommand{\Sussex}{Department of Physics and Astronomy, University of 
Sussex, Falmer, Brighton BN1 9QH, United Kingdom}
\newcommand{\Syracuse}{Department of Physics, Syracuse University,
Syracuse NY 13210, USA}
\newcommand{\Tennessee}{Department of Physics and Astronomy, 
University of Tennessee, Knoxville, Tennessee 37996, USA}
\newcommand{\Texas}{Department of Physics, University of Texas at Austin, 
Austin, Texas 78712, USA}
\newcommand{\Tufts}{Department of Physics and Astronomy, Tufts University, Medford, 
Massachusetts 02155, USA}
\newcommand{\UCL}{Physics and Astronomy Department, University College 
London, 
Gower Street, London WC1E 6BT, United Kingdom}
\newcommand{\Virginia}{Department of Physics, University of Virginia, 
Charlottesville, Virginia 22904, USA}
\newcommand{\WSU}{Department of Mathematics, Statistics, and Physics,
 Wichita State University, 
Wichita, Kansas 67260, USA}
\newcommand{\WandM}{Department of Physics, William \& Mary, 
Williamsburg, Virginia 23187, USA}
\newcommand{\Wisconsin}{Department of Physics, University of 
Wisconsin-Madison, Madison, Wisconsin 53706, USA}
\affiliation{\ANL}
\affiliation{\Atlantico}
\affiliation{\Bandirma}
\affiliation{\BHU}
\affiliation{\Caltech}
\affiliation{\CUSB}
\affiliation{\Charles}
\affiliation{\Cincinnati}
\affiliation{\CSU}
\affiliation{\CTU}
\affiliation{\Delhi}
\affiliation{\Erciyes}
\affiliation{\FNAL}
\affiliation{\FSU}
\affiliation{\UFG}
\affiliation{\Guwahati}
\affiliation{\Homi}
\affiliation{\Houston}
\affiliation{\Hyderabad}
\affiliation{\IHyderabad}
\affiliation{\IIT}
\affiliation{\Imperial}
\affiliation{\Indiana}
\affiliation{\ICS}
\affiliation{\INR}
\affiliation{\IOP}
\affiliation{\UIowa}
\affiliation{\ISU}
\affiliation{\Irvine}
\affiliation{\JINR}
\affiliation{\Magdalena}
\affiliation{\MSU}
\affiliation{\Duluth}
\affiliation{\Minnesota}
\affiliation{\Mississippi}
\affiliation{\NISER}
\affiliation{\OSU}
\affiliation{\Panjab}
\affiliation{\Pitt}
\affiliation{\QMU}
\affiliation{\SAlabama}
\affiliation{\Carolina}
\affiliation{\SMU}
\affiliation{\Sussex}
\affiliation{\Syracuse}
\affiliation{\Texas}
\affiliation{\Tufts}
\affiliation{\UCL}
\affiliation{\Virginia}
\affiliation{\WSU}
\affiliation{\WandM}
\affiliation{\Wisconsin}

\author{S.~Abubakar}
\affiliation{\Erciyes}

\author{M.~A.~Acero}
\affiliation{\Atlantico}

\author{B.~Acharya}
\affiliation{\Mississippi}

\author{P.~Adamson}
\affiliation{\FNAL}









\author{N.~Anfimov}
\affiliation{\JINR}


\author{A.~Antoshkin}
\affiliation{\JINR}


\author{E.~Arrieta-Diaz}
\affiliation{\Magdalena}

\author{L.~Asquith}
\affiliation{\Sussex}


\author{A.~Aurisano}
\affiliation{\Cincinnati}






\author{N.~Balashov}
\affiliation{\JINR}

% ...

\collaboration{The NOvA Collaboration}
\noaffiliation




\date{\today}

\begin{abstract}
We report a search for neutrino oscillations to sterile neutrinos in the NOvA detectors under a model with three active and one sterile neutrinos.  This search simultaneously fits data in the two NOvA detectors and is the first from NOvA to use both neutrino- and antineutrino-mode beams, with exposures of $26.61\times10^{20}$ and $12.50\times10^{20}$ protons on target, respectively. There is no evidence for sterile neutrinos in the data and we are able to exclude regions of parameter space that were allowed by previous experiments, including most of the allowed region reported by IceCube.
\end{abstract}

\maketitle


The phenomenon of mixing among three neutrino mass states, $\nu_{1}$, $\nu_{2}$, $\nu_{3}$, and three neutrino flavor states, $\nu_{\mu}$, $\nu_{e}$, $\nu_{\tau}$, is well established~\cite{Super-Kamiokande:1998kpq,SNO:2002tuh,KamLAND:2004mhv,K2K:2006yov,MINOS:2006foh,T2K:2011ypd,DayaBay:2012fng,RENO:2012mkc,DoubleChooz:2014kuw,OPERA:2015wbl,NOvA:2016kwd}. This three-flavor (3F) mixing is described by three mixing angles, $\theta_{12}$, $\theta_{13}$ and $\theta_{23}$, which determine the amplitude of the mixing; two mass-squared splittings, $\Delta m^{2}_{21}$ and $\Delta m^{2}_{32}$, where $\Delta m_{ji}^{2} \equiv m_{j}^{2} - m_{i}^{2}$, which determine the frequency of the mixing for a given neutrino energy, $E_{\nu}$, and distance traveled by the neutrino, or baseline, $L$; and a {\it CP} violating phase, $\delta_{CP}$, which allows for differences in mixing among neutrinos when compared to antineutrinos.  The 3F mixing successfully describes the data from a wide variety of experiments studying different neutrino flavors, energies and baselines. Moreover, the LEP measurements of the $Z$ boson decay width show that only three light neutrino species couple to the $Z$~\cite{ALEPH:2005ab}.

There have been a number of anomalous results reported~\cite{LSND:2001aii,MiniBooNE:2020pnu,Barinov:2022wfh,SAGE:1998fvr,GALLEX:1997lja,Huber:2011wv} that, if they were the result of neutrino mixing, would require at least one more neutrino mass state and flavor state. Any additional state must not interact via the weak force as it does not couple to the $Z$ and would be deemed sterile. The anomalous results tend to suggest an additional mass-squared splitting of $\Delta m^2 \sim \qty{1}{eV^2} \gg \Delta m^2_{32}, \Delta m^2_{21}$, but are in tension with results from many experiments that have found no evidence for sterile neutrinos with mass-squared splittings from 10$^{-3}$~eV$^{2}$ -- 10$^{2}$~eV$^{2}$~\cite{KARMEN:2002zcm,MicroBooNE:2021tya,MINOS:2020iqj,Dydak:1983zq, Stockdale:1984cg, T2K:2019efw, SciBooNE:2011qyf, Super-Kamiokande:2014ndf, MINOS:2010fgd, MINOS:2016viw, MINOS:2020iqj,  CCFRNuTeV:1998gjj, FERMILABE531:1986uhg, CHORUS:2007wlo, NOMAD:2001xxt,OPERA:2019kzo,IceCube:2024dlz,NOvA:2024imi}.

The simplest extension to 3F mixing that would accommodate the additional mass-squared splitting is the 3+1 model that adds one sterile neutrino to the known neutrino flavor states~\cite{Caldwell:1993kn, Peltoniemi:1993ec, Bilenky:1998dt, Barger:2000ch, Goldman:2000sc}. This model introduces six new parameters: $\Delta m^{2}_{41}$, $\theta_{14}$, $\theta_{24}$, $\theta_{34}$, $\delta_{2}$ and $\delta_{3}$~\cite{MINOS:2010fgd}. Under this model, the active neutrino survival probability can be approximated as \cite{MINOS:2016viw}
\begin{equation}\label{eq:prob_nc}
    \begin{aligned}
    1 - P(\overset{\scriptscriptstyle(-)}{\nu}_{\mu} \rightarrow \nu_{s}) \approx 1 & - \cos^{4}\theta_{14}\cos^{2}\theta_{34}\sin^{2}2\theta_{24}\sin^{2}\Delta_{41} \\
    & - \sin^{2}\theta_{34}\sin^{2}2\theta_{23}\sin^{2}\Delta_{31} \\
    & \overset{(-)}{+} \frac{1}{2}\sin\delta_{2}\sin\theta_{24}\sin2\theta_{23}\sin\Delta_{31},
    \end{aligned}
\end{equation}
and the $\overset{\scriptscriptstyle(-)}{\nu}_{\mu}$ survival probability can be approximated as
\begin{equation}\label{eq:prob_cc_numu}
    \begin{aligned}
    P(\overset{\scriptscriptstyle(-)}{\nu}_{\mu} \rightarrow \overset{\scriptscriptstyle(-)}{\nu}_{\mu}) \approx 1 & - \sin^{2}2\theta_{24}\sin^{2}\Delta_{41} \\
    & + 2\sin^{2}2\theta_{23}\sin^{2}\theta_{24}\sin^{2}\Delta_{31} \\
    & - \sin^{2}2\theta_{23}\sin^{2}\Delta_{31},
    \end{aligned}
\end{equation}
where $\Delta_{ji} \equiv \frac{\Delta m_{ji}^{2} L}{4E_{\nu}}$. Exact oscillation probability calculations are used for the analysis described below.

Equations~\eqref{eq:prob_nc} and~\eqref{eq:prob_cc_numu} show that sterile neutrinos modify the magnitude of oscillations at the atmospheric frequency, $\Delta m^2_{31}$, and introduce a new sterile frequency driven by $\Delta m^2_{41}$. At baselines greater than $\sim$100~km, the frequency of sterile oscillations at $\Delta m_{41}^{2} > \qty{0.05}{\eV^2}$ is too high for most experiments to resolve. These oscillations manifest as a downward normalization shift in the neutrino energy spectrum. At shorter baselines of $\sim$1~km, energy-dependent sterile oscillations arise when $\Delta m_{41}^{2} > \qty{0.5}{\eV^2}$, as seen in Fig.~1 of~\cite{NOvA:2024imi}. Fitting data from both baseline regimes simultaneously extends the region of parameter space to which an experiment is sensitive. Additionally, incorporating information from the $\nu_{\mu}$ charged current (CC) and neutral current (NC) channels makes an analysis sensitive to the atmospheric 3F oscillation parameters, $\theta_{23}$ and $\Delta m_{32}^{2}$, as well as to the sterile-related parameters $\theta_{24}$, $\theta_{34}$, $\Delta m_{41}^{2}$, and $\delta_{24}$. Sensitivity to $\theta_{14}$ and $\delta_{14}$ relies on the inclusion of a $\nu_{e}$ appearance sample that is outside the scope of this Letter.

We analyzed data from the NOvA experiment to search for potential mixing between active and sterile neutrinos, using a simultaneous fit of CC and NC data from both the NOvA Near (ND) and Far Detectors (FD), in both neutrino and antineutrino beam modes.  These detectors are functionally identical low-$Z$ detectors, with the ND at a distance of 1~km from the beam target, and the FD at a distance of 810 km. The ND is located 100 meters underground at Fermilab, an overburden equivalent to 225 meters of water. The FD is located in northern Minnesota with a modest overburden equivalent to 3 meters of water; the cosmic ray rate of  130~kHz is the primary source of background events there. Both detectors are centered 14.6 mrad from the axis of Fermilab’s Neutrinos from the Main Injector (NuMI) beam~\cite{Adamson:2015dkw}. 

The detectors consist of planes of 6.6~cm by 3.9~cm PVC cells containing liquid scintillator~\cite{Mufson:2015kga} which alternate between vertical and horizontal orientations. The \qty{15.5}{m} (\qty{3.9}{m}) by \qty{15.5}{m} (\qty{3.9}{m}) by \qty{60}{m} (\qty{13}{m})  FD (ND) has a mass of \qty{14}{kT} (\qty{0.3}{kT}), of which 62\% is liquid scintillator.  The ND includes a \qty{3}{m} long muon range stack at the downstream end that interleaves \qty{10}{cm} thick steel plates  with pairs of horizontal and vertical planes in the region. Wavelength-shifting fibers collect the light produced by charged particles traversing each cell. The light from each cell is read out by a pixel of an avalanche photodiode and digitized using custom readout electronics with nanosecond timing resolution. Signals meeting a minimum pulse height requirement within a \qty{550}{\micro\s} window around the beam pulse are saved for offline analysis. The FD cosmic background is sampled by a 
minimum bias trigger~\cite{NOvA:2020dll}.

Fermilab’s Main Injector accelerator delivers \qty{120}{GeV} protons to a graphite target approximately every \qty{1.3}{s}. The pions and kaons produced in the resulting interactions are focused by two magnetic horns; the current delivered to the horns can be adjusted to produce a beam of either predominantly $\nu_{\mu}$ or $\bar{\nu}_{\mu}$ that result from the hadron decays. In the range \qty{1}{GeV} -- \qty{5}{GeV} the $\nu$-mode ($\overline{\nu}$-mode) beam is 94\% (93\%) $\nu_{\mu}$ ($\bar{\nu}_{\mu}$), with the largest contaminant being $\bar{\nu}_{\mu}$ ($\nu_{\mu}$). The contamination from $\nu_{e}$ and $\bar{\nu}_{e}$ is less than 1\% in either mode.  The off-axis location of the NOvA detectors results in a narrow beam peaked at \qty{2}{GeV} with a suppressed high-energy tail. 

We simulate the neutrino beam flux, neutrino interactions and particle propagation in the detector using the Geant4~\cite{GEANT4:2002zbu} and GENIE~\cite{Andreopoulos:2015wxa} software packages. The detector response is simulated with NOvA-specific routines. The details of the simulation can be found in~\cite{NOvA:2022wnj}.

This analysis builds on previous NOvA searches for active-sterile neutrino mixing~\cite{NOvA:2024imi} by using both $\nu$-mode and $\bar{\nu}$-mode data. The $\nu$-mode sample corresponds to $26.61\times10^{20}$ protons delivered to the target (POT), or twice the data previously presented, while the newly added $\bar{\nu}$-mode sample includes $12.50\times10^{20}$~POT. The data were recorded by the FD between February 6, 2014 and March 14, 2024, with a beam-on time of \qty{885.53}{s} for $\nu$-mode and \qty{322.59}{s} for $\bar{\nu}$-mode. This analysis represents the first simultaneous fit of the NOvA neutrino and antineutrino data across both detectors in a sterile neutrino search. Additionally, this is the first NOvA search allowing an extensive range of $\Delta m_{41}^2$ values with antineutrino data rather than a single constrained value~\cite{NOvA:2021smv}.

The analysis uses samples of $\nu_{\mu}$ and $\bar{\nu}_{\mu}$ CC interactions and NC interactions from both the ND and FD.  The CC candidates have a muon in the final state while the NC candidates only contain hadronic activity from the transfer of energy and momentum by the (anti)neutrino to the nucleus but no final state lepton. The CC candidates are selected according to the requirements used by the 3F analysis described in~\cite{NOvA:2021nfi}. The NC candidates are selected from the interactions that have not been identified as either $\nu_{\mu}$~$(\overline{\nu}_{\mu})$~CC or $\nu_{e}$~$(\overline{\nu}_{e})$~CC.
\begin{figure}[!ht]
      \includegraphics[height=0.7\textheight]
      {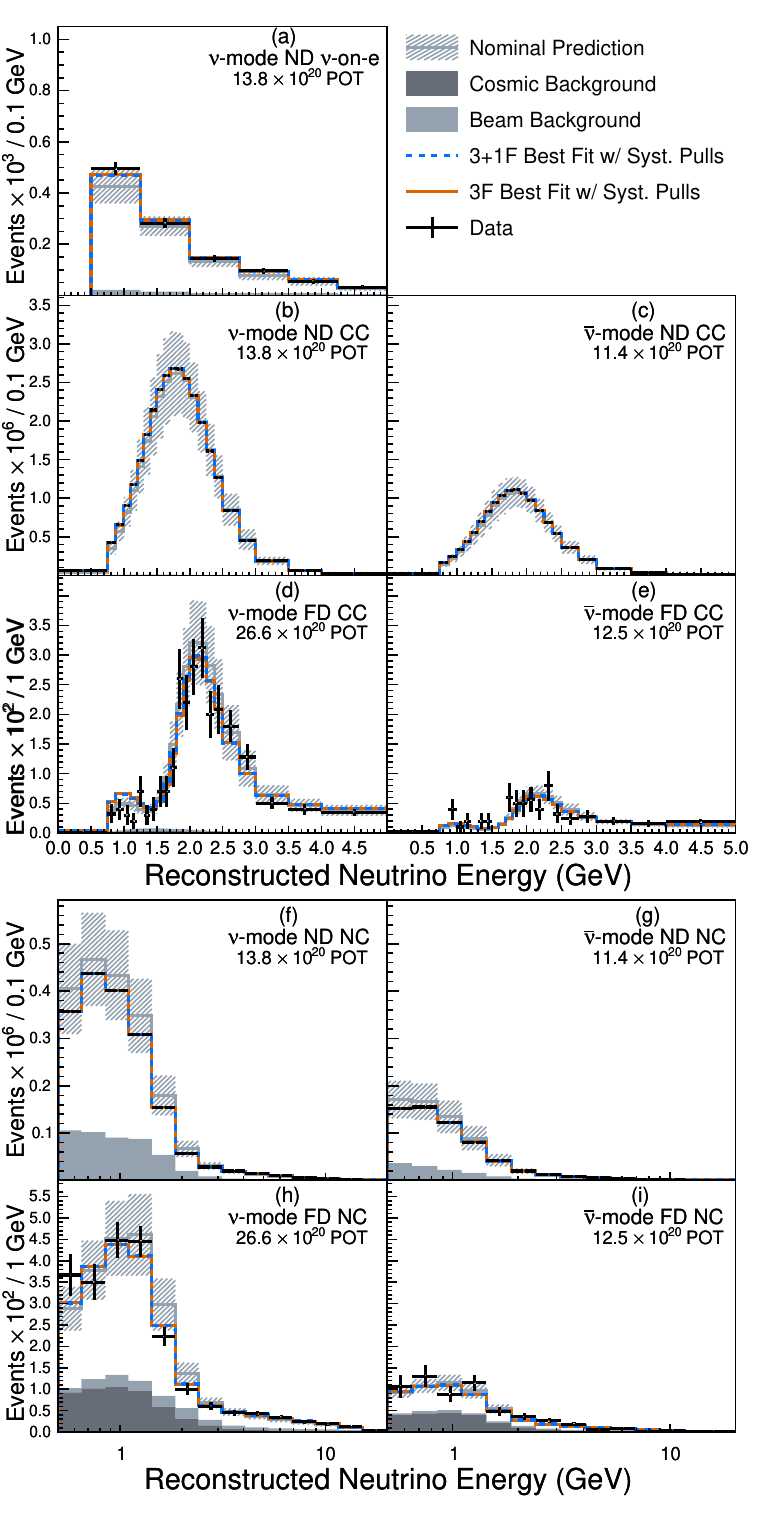}
    \caption{Reconstructed neutrino energy spectra for the selected candidates.  The type of candidate, detector and beam configuration for each spectrum is indicated in its panel and the backgrounds are shown as a stacked histogram. 
    The 3F expectation including systematic uncertainties shown in orange assumes $\Delta m^2_{32} = 2.51 \times 10^{-3}~\mathrm{eV}^{2}$ and $\theta_{23} = \ang{49.6}$. 
    The best fit is the dotted blue line.}
    \label{fig:spectra}
\end{figure}

The potential NC interaction candidates are required to have a reconstructed vertex within the fiducial volume indicated in Table~\ref{tab:fiducial}, at least \num{1} reconstructed particle, all reconstructed particles must be contained within the volume indicated in the table, and the candidate must cross at least \num{3} contiguous planes in the detector. 
\begin{table}[h]
\caption{Reconstructed vertex and particle containment requirements for candidate NC interactions in each detector. Values shown are centimeters from the indicated detector face. In the ND, the rear face is the start of the muon range stack.}
\label{tab:fiducial}
\begin{tabular}{l|l|c|c|c|c|c|c}
\hline \hline
Detector & Category & Front & Rear & Top & Bottom & East & West  \\ \hline
\multirow{2}{*}{ND} & Vertex & 300 & 260 & 120 & 80 & 120 & 80 \\ 
 & Containment & 150 & 50 & 20 & 20 & 20 & 20 \\ \hline
\multirow{2}{*}{FD} & Vertex and & \multirow{2}{*}{160} & \multirow{2}{*}{160} & \multirow{2}{*}{100} & \multirow{2}{*}{100} & \multirow{2}{*}{100} & \multirow{2}{*}{100} \\ 
 & Containment & & & & & &  \\ \hline
\hline
\end{tabular}
\end{table}
The asymmetric vertex requirements in the ND exclude locations closer to the beam axis, which are more likely to have activity from particles generated in the rock near the detector. 
Similarly, the ND particle containment requirements are more stringent at the front and rear faces of the detector compared with the requirements on the other faces to reject background candidates due to interactions in rock upstream of the detector and CC beam interactions, respectively. The larger size of the FD and the lower rate of interactions there allow for more symmetric vertex and containment requirements.
The shallower overburden of the FD results in a cosmic background interaction that is significantly higher than at the ND. We use information about the reconstructed vertex, shower shapes and energy, number of hits and the transverse momentum fraction to construct a boosted decision tree to reject background cosmic rays. 

A convolutional neural network (CNN)~\cite{Aurisano:2016jvx,Psihas:2019ksa} identifies the NC interactions in the sample surviving the vertex, containment and cosmic ray selections. The CNN, which is trained separately for $\nu$-mode and $\bar{\nu}$-mode data, provides probability scores for different neutrino interaction types based on energy depositions in the detector. The optimal requirements are determined by using a figure of merit that considers the systematic and statistical uncertainties on the samples. In the FD, the CNN and cosmic rejection requirements are jointly tuned. 

The deposited energy of NC interaction candidates is estimated by taking a weighted sum of the hadronic and electromagnetic components of the calorimetric energy in the detector. An additional bias correction is applied as a function of total calorimetric energy. The overall NC energy resolution is 40\% in the ND and 30\% in the FD for both beam configurations. The resolution is better in the FD as more of the energy of the interaction is contained in its larger volume. The event selection criteria and neutrino energy estimator used for the $\nu_{\mu}$ CC samples are described in \cite{NOvA:2021nfi}.

Figure~\ref{fig:spectra} shows the reconstructed energy spectra for the samples used in this analysis.
The numbers of candidate CC and NC interactions selected in the data and simulation, assuming 3F oscillations using the values $\Delta m^2_{32} = 2.51 \times 10^{-3}~\mathrm{eV^2}$ and $\theta_{23} = \qty{49.6}{\degree}$ favored by a global fit to the experimental data~\cite{Esteban:2020cvm}, for each detector and beam configuration are shown in Table~\ref{tab:selected}. 

\begin{table}[h]
\caption{Number of selected candidate interactions. The letters in parentheses indicate the corresponding panel in Fig.~\ref{fig:spectra}. The neutrino-on-electron ($\nu$-on-e) sample is used to constrain uncertainties on the beam flux as described in the text.}
\label{tab:selected}
\begin{tabular}{l@{\hspace{0.5em}}|l@{\hspace{0.5em}}|l|l|l}
\hline \hline
\multicolumn{3}{c|}{\multirow{2}{*}{Category}} & Selected & Predicted          \\
\multicolumn{3}{c|}{}                          & Number   & Number             \\ \hline
\multirow{6}{*}{CC}&\multirow{3}{*}{ND} & \T $\nu$-on-e~~\thinspace(a)       & $827$              & $747.99\pm99.85$          \\
                      & & $\nu$-mode (b)          & $3.59\times10^{6}$ & $3.51\pm0.65\times10^{6}$ \\
                      & & \B $\bar{\nu}$-mode (c) & $1.47\times10^{6}$ & $1.45\pm0.24\times10^{6}$ \\
&\multirow{2}{*}{FD} & \T $\nu$-mode (d)       & 384                & $377.37\pm71.12$          \\
                      & & \B $\bar{\nu}$-mode (e) & 106                & $91.02\pm15.02$           \\ \hline
\multirow{5}{*}{NC} &\multirow{2}{*}{ND}& \T $\nu$-mode (f)       & $5.47\times10^{5}$ & $5.78\pm1.08\times10^{5}$ \\
                       & & \B $\bar{\nu}$-mode (g) & $1.81\times10^{5}$ & $1.80\pm0.41\times10^{5}$ \\ 
&\multirow{2}{*}{FD} & \T $\nu$-mode (h)       & 889                & $917.87\pm144.72$         \\
                       & & \B $\bar{\nu}$-mode (i) & 222                & $205.16\pm31.22$          \\ \hline \hline
\end{tabular}
\end{table}

A sterile neutrino could manifest across several orders of magnitude in $\Delta m^{2}_{41}$ meaning there could be evidence at either the baseline of the ND or FD.  Thus, the current analysis cannot constrain neutrino cross-section models using the ND data prior to fitting the FD data, nor can those data be used to extrapolate an expected number of interaction candidates at the FD. Instead, this analysis simultaneously fits the neutrino and antineutrino data in the ND and FD.

We consider systematic uncertainties on the detector modeling, neutrino interactions, and beam flux as described in~\cite{NOvA:2021nfi}, with three notable differences. The first is that due to the potential for sterile neutrinos to impact the observed ND data, this analysis uses the GENIE~3.0.6 simulation~\cite{Andreopoulos:2009rq,Andreopoulos:2015wxa} without changes. The meson exchange current (MEC) component of the simulation is the least well understood and we assessed the uncertainty due to this component by developing spectral shape and normalization uncertainties based on the model spread of  the Val\`encia \cite{Nieves:2011pp}, SuSA \cite{Megias:2016fjk}, and GENIE empirical \cite{Katori:2013eoa} MEC models. The second difference is the treatment of the uncertainty in the kaon component of the secondary beam. These kaons are the progenitors of many candidate NC interactions selected for this analysis. The previous sterile neutrino search using $\nu$-mode data~\cite{NOvA:2024imi} determined the uncertainty on these NC progenitors to be 10\%; the previous $\bar{\nu}$-mode search~\cite{NOvA:2021smv} used a higher 30\% uncertainty on the same component of NC progenitors in that sample. We use the same uncertainties for these samples in this analysis. The third difference is the addition of candidate neutrino-on-electron ($\nu$-on-$e$) scattering interactions in the ND from the $\nu$-mode beam. That cross section is well-known and the inclusion of the sample allows for the constraint of flux-related systematic uncertainties~\cite{Oh:2025cpz}. Table~\ref{tab:selected} shows the number of selected and expected candidate interactions for this sample. 

For each source of systematic uncertainty we randomly vary model parameters within their uncertainties to produce a new systematically-fluctuated ``universe,'' $u$. We then construct a matrix encoding the covariance between each bin of reconstructed energy, $i$ and $j$, and each oscillation channel, $\alpha$ and $\beta$,
\begin{equation}
    C^{u}_{\alpha i,\beta j} = \frac{1}{N^{\mathrm{CV}}_{\alpha i}N^{\mathrm{CV}}_{\beta j}}[N^{\mathrm{CV}}_{\alpha i}-N^{u}_{\alpha i}]\times[N^{\mathrm{CV}}_{\beta j}-N^{u}_{\beta j}],
    \label{eq:covariance}
\end{equation}
where $N^{\mathrm{CV}}_{\alpha i(\beta j)}$ represents the number of events in the $i$th ($j$th) reconstructed energy bin from Fig.~\ref{fig:spectra} for the nominal simulation in the specified oscillation channels and $N^{u}_{\alpha i(\beta j)}$ is the number of events in the bin for the systematically varied universe. The covariance defined above is the fractional covariance between the bins, which, when separated by oscillation channel, is insensitive to different combinations of oscillation parameters during the fit. We sum the fractional covariances found from Eq.~\eqref{eq:covariance} for all universes and divide the sum by the total number of universes to get the average fractional covariance. Including the oscillation channels in the covariance matrix makes the matrix independent of the combination of oscillation parameters. The total covariance matrix is the sum of the matrices for each systematic uncertainty.

We combine a Poisson log-likelihood treatment of statistical uncertainties with a Gaussian multivariate treatment of systematic uncertainties. The covariance matrix described above is used to determine optimal systematic weights, $s_{\alpha i}$, for each oscillation channel and analysis bin. The test statistic is
\begin{equation}
    \begin{aligned}
    \chi^{2} = 2 &\sum_{i} \left[ S_{i} - O_{i} + O_{i} \log \left( \frac{O_{i}}{S_{i}} \right) \right] \\
    + &\sum_{ij} \sum_{\alpha \beta} (s_{\alpha i} - 1) C_{\alpha i \beta j}^{-1} (s_{\beta j} - 1),
    \end{aligned}
\end{equation}
where $O_{i}$ are the observed data  and $S_{i} = \sum_{\alpha} s_{\alpha i} p_{\alpha i}$ are the weighted predictions accounting for both the systematic uncertainties, $s_{\alpha i}$, and oscillations, $p_{\alpha i}$.

\begin{figure*}[!ht]
    \centering
    \subfloat{
      \includegraphics[width=0.9\columnwidth]{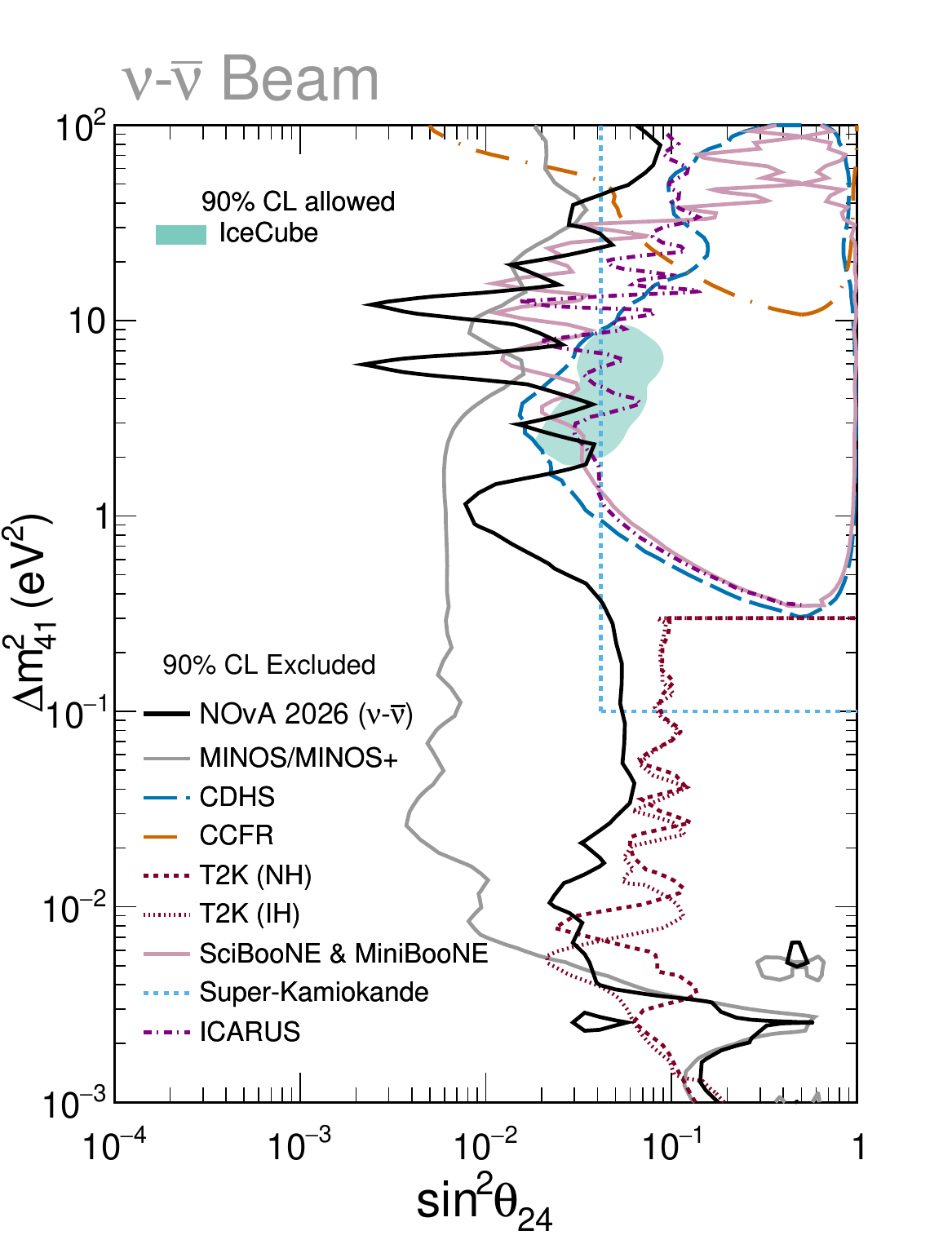}
    }
    \hspace{1em}
    \subfloat{
      \includegraphics[width=0.9\columnwidth]{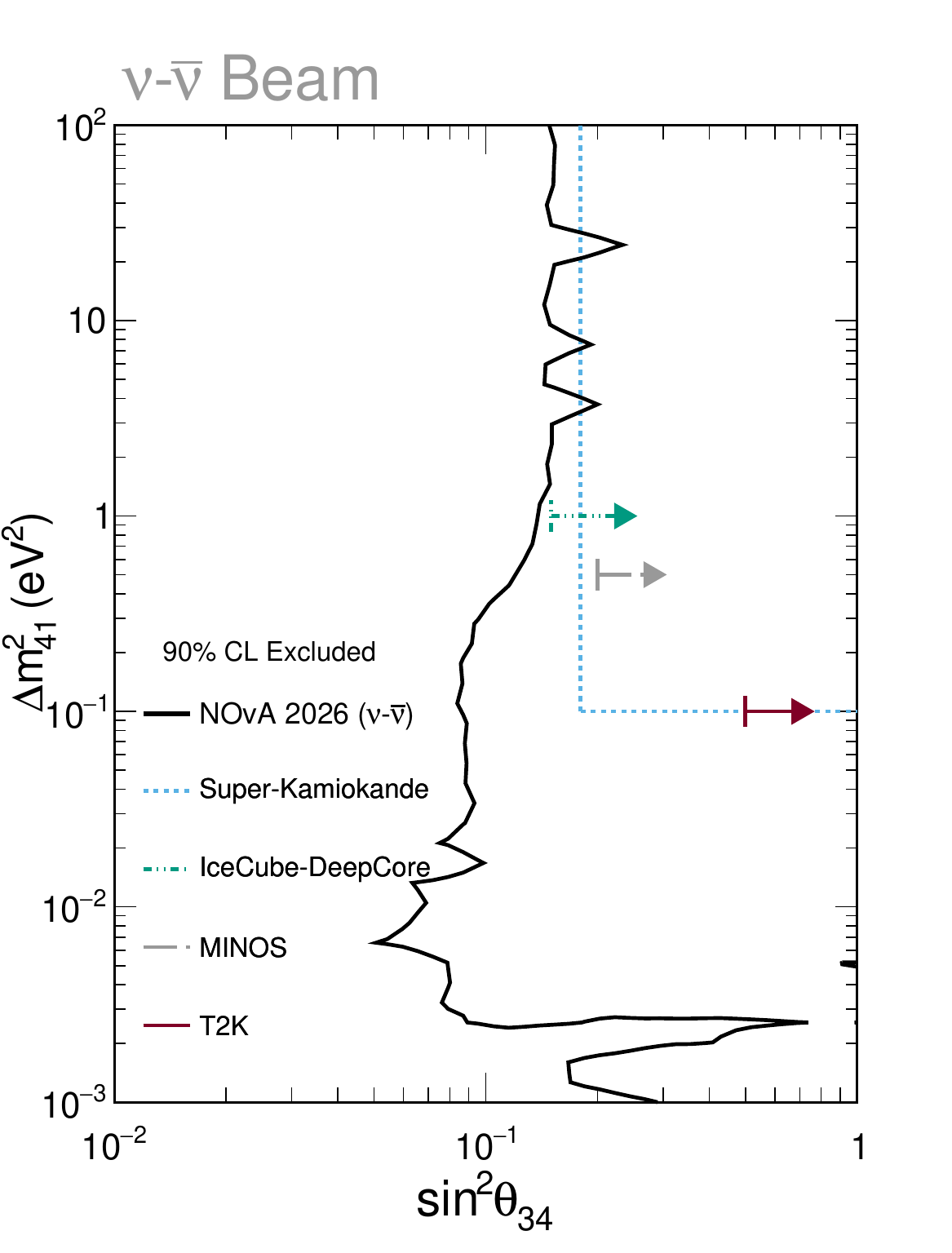}%
    }
    \caption{NOvA's 90\% confidence limits in $\Delta m_{41}^{2}-\sin^{2}\theta_{24}$ space (left), and $\Delta m_{41}^{2}-\sin^{2}\theta_{34}$ space (right), with allowed regions and exclusion contours from other experiments. Regions to the right of open contours are excluded; the small rejected and allowed regions at low $\Delta m_{41}^{2}$ are explained in the text.
    }
    \label{fig:contours}
\end{figure*}

In this analysis the 3F atmospheric oscillation parameter $\theta_{23}$ varies freely in the fit. The mass-squared splitting central value of $\Delta m_{32}^{2} = |2.51\times 10^{-3}|$~eV$^{2}$ is used for both orderings. The fit to the mass-squared splitting is constrained to be within $\pm0.15\times10^{-3}$~eV$^{2}$ of the central value to pin the fit to a $3+1$ flavor paradigm; this constraint is double the Gaussian $3\sigma$ range derived from a 2020 global fit to data including atmospheric neutrino oscillations~\cite{Esteban:2020cvm}. The sterile parameters $\Delta m_{41}^{2}$, $\theta_{24}$, $\theta_{34}$ and $\delta_{2}$ freely vary during the fit. The remaining sterile parameters are held fixed at \num{0} in the fit, due to constraints from solar and reactor experiments \cite{DayaBay:2022orm} and unitarity \cite{Parke:2015goa}. The results indicate no evidence of sterile neutrinos and are consistent with 3F oscillations at 90\% confidence~\cite{Chaudhary:2026gtx} as seen in Fig.~\ref{fig:contours}. The confidence intervals shown in the figure were produced using the Profiled Feldman-Cousins technique~\cite{NOvA:2022wnj}.

Our $\Delta m^{2}_{41}$-$\sin^{2}(\theta_{34})$ space limits are the best limits to date, with $\sin^{2}(\theta_{34})\gtrsim0.2$ excluded across nearly five orders of magnitude in $\Delta m^{2}_{41}$. Moreover, this analysis excludes more parameter space for combinations of $\Delta m^{2}_{41}\gtrsim4$~eV$^{2}$ and $\sin^{2}(\theta_{24})\lesssim0.02$ than previous results~\cite{IceCube:2020phf, MINOS:2020iqj, Dydak:1983zq, Stockdale:1984cg, T2K:2019efw, SciBooNE:2011qyf, Super-Kamiokande:2014ndf, MINOS:2016viw, IceCube:2017ivd, CCFRNuTeV:1998gjj, FERMILABE531:1986uhg, CHORUS:2007wlo, NOMAD:2001xxt,OPERA:2019kzo}, including nearly all parameter space allowed by IceCube at 90\% confidence level~\cite{IceCubeCollaboration:2024nle}.  The small region centered near $\Delta m^{2}_{41}\sim2\times10^{-3}$~eV$^{2}$ and $\sin^{2}(\theta_{24})\sim0.04$ is rejected because the simultaneous fit to the $\nu$-mode and $\overline{\nu}$-mode NC samples tightly constrains $\delta_{24}$ to be near 0 in that region of parameter space; there is not enough freedom in the mixing angles $\theta_{23}$ and $\theta_{34}$ to achieve a good fit across all data samples for that region of parameter space. The small allowed region near $\Delta m^{2}_{41}\sim5\times10^{-3}$~eV$^{2}$ and $\sin^{2}(\theta_{24})\sim0.5$ is due to degeneracies among the atmospheric and potential sterile neutrino mixing parameters.  

In summary, we have presented the first search for sterile neutrino oscillations under the 3+1 oscillation paradigm using a simultaneous fit to $\nu$-mode and $\bar{\nu}$-mode data in the NOvA ND and FD. We find no evidence for sterile neutrinos and exclude extensive regions of $\Delta m_{41}^{2}-\sin^{2}\theta_{34}$ parameter space and new regions of $\Delta m_{41}^{2}-\sin^{2}\theta_{24}$ parameter space, including most of the IceCube-allowed region. 

This document was prepared by the NOvA collaboration using the resources of the Fermi National Accelerator Laboratory (Fermilab), a U.S. Department of Energy, Office of Science, HEP User Facility. Fermilab is managed by Fermi Forward Discovery Group, LLC, acting under Contract No. 89243024CSC000002.  This work was supported by the U.S. Department of Energy; the U.S. National Science Foundation; the Department of Science and Technology, India; the European Research Council; the MSMT CR, GA UK, Czech Republic; the RAS, the Ministry of Science and Higher Education, and RFBR, Russia; CNPq and FAPEG, Brazil; UKRI, STFC and the Royal Society, United Kingdom; and the state and University of Minnesota.  We are grateful for the contributions of the staffs of the University of Minnesota at the Ash River Laboratory, and of Fermilab. For the purpose of open access, the author has applied a Creative Commons Attribution (CC BY) license to any Author Accepted Manuscript version arising.

\bibliographystyle{apsrev4-2}
\bibliography{bibliography}

@article{Super-Kamiokande:1998kpq,
    author = "Fukuda, Y. and others",
    collaboration = "Super-Kamiokande",
    title = "{Evidence for oscillation of atmospheric neutrinos}",
    eprint = "hep-ex/9807003",
    archivePrefix = "arXiv",
    reportNumber = "BU-98-17, ICRR-REPORT-422-98-18, UCI-98-8, KEK-PREPRINT-98-95, LSU-HEPA-5-98, UMD-98-003, SBHEP-98-5, TKU-PAP-98-06, TIT-HPE-98-09",
    doi = "10.1103/PhysRevLett.81.1562",
    journal = "Phys. Rev. Lett.",
    volume = "81",
    pages = "1562--1567",
    year = "1998"
}

@article{SNO:2002tuh,
    author = "Ahmad, Q. R. and others",
    collaboration = "SNO",
    title = "{Direct evidence for neutrino flavor transformation from neutral current interactions in the Sudbury Neutrino Observatory}",
    eprint = "nucl-ex/0204008",
    archivePrefix = "arXiv",
    doi = "10.1103/PhysRevLett.89.011301",
    journal = "Phys. Rev. Lett.",
    volume = "89",
    pages = "011301",
    year = "2002"
}

@article{KamLAND:2004mhv,
    author = "Araki, T. and others",
    collaboration = "KamLAND",
    title = "{Measurement of neutrino oscillation with KamLAND: Evidence of spectral distortion}",
    eprint = "hep-ex/0406035",
    archivePrefix = "arXiv",
    doi = "10.1103/PhysRevLett.94.081801",
    journal = "Phys. Rev. Lett.",
    volume = "94",
    pages = "081801",
    year = "2005"
}

@article{K2K:2006yov,
    author = "Ahn, M. H. and others",
    collaboration = "K2K",
    title = "{Measurement of Neutrino Oscillation by the K2K Experiment}",
    eprint = "hep-ex/0606032",
    archivePrefix = "arXiv",
    doi = "10.1103/PhysRevD.74.072003",
    journal = "Phys. Rev. D",
    volume = "74",
    pages = "072003",
    year = "2006"
}

@article{MINOS:2006foh,
    author = "Michael, D. G. and others",
    collaboration = "MINOS",
    title = "{Observation of muon neutrino disappearance with the MINOS detectors and the NuMI neutrino beam}",
    eprint = "hep-ex/0607088",
    archivePrefix = "arXiv",
    reportNumber = "FERMILAB-PUB-06-243, BNL-76806-2006-JA",
    doi = "10.1103/PhysRevLett.97.191801",
    journal = "Phys. Rev. Lett.",
    volume = "97",
    pages = "191801",
    year = "2006"
}

@article{T2K:2011ypd,
    author = "Abe, K. and others",
    collaboration = "T2K",
    title = "{Indication of Electron Neutrino Appearance from an Accelerator-produced Off-axis Muon Neutrino Beam}",
    eprint = "1106.2822",
    archivePrefix = "arXiv",
    primaryClass = "hep-ex",
    doi = "10.1103/PhysRevLett.107.041801",
    journal = "Phys. Rev. Lett.",
    volume = "107",
    pages = "041801",
    year = "2011"
}

@article{DayaBay:2012fng,
    author = "An, F. P. and others",
    collaboration = "Daya Bay",
    title = "{Observation of electron-antineutrino disappearance at Daya Bay}",
    eprint = "1203.1669",
    archivePrefix = "arXiv",
    primaryClass = "hep-ex",
    doi = "10.1103/PhysRevLett.108.171803",
    journal = "Phys. Rev. Lett.",
    volume = "108",
    pages = "171803",
    year = "2012"
}

@article{RENO:2012mkc,
    author = "Ahn, J. K. and others",
    collaboration = "RENO",
    title = "{Observation of Reactor Electron Antineutrino Disappearance in the RENO Experiment}",
    eprint = "1204.0626",
    archivePrefix = "arXiv",
    primaryClass = "hep-ex",
    doi = "10.1103/PhysRevLett.108.191802",
    journal = "Phys. Rev. Lett.",
    volume = "108",
    pages = "191802",
    year = "2012"
}

@article{DoubleChooz:2014kuw,
    author = "Abe, Y. and others",
    collaboration = "Double Chooz",
    title = "{Improved measurements of the neutrino mixing angle $\theta_{13}$ with the Double Chooz detector}",
    eprint = "1406.7763",
    archivePrefix = "arXiv",
    primaryClass = "hep-ex",
    doi = "10.1007/JHEP02(2015)074",
    journal = "JHEP",
    volume = "10",
    pages = "086",
    year = "2014",
    note = "[Erratum: JHEP 02, 074 (2015)]"
}

@article{OPERA:2015wbl,
    author = "Agafonova, N. and others",
    collaboration = "OPERA",
    title = "{Discovery of $\tau$ Neutrino Appearance in the CNGS Neutrino Beam with the OPERA Experiment}",
    eprint = "1507.01417",
    archivePrefix = "arXiv",
    primaryClass = "hep-ex",
    doi = "10.1103/PhysRevLett.115.121802",
    journal = "Phys. Rev. Lett.",
    volume = "115",
    number = "12",
    pages = "121802",
    year = "2015"
}

@article{NOvA:2016kwd,
    author = "Adamson, P. and others",
    collaboration = "NOvA",
    title = "{First measurement of electron neutrino appearance in NOvA}",
    eprint = "1601.05022",
    archivePrefix = "arXiv",
    primaryClass = "hep-ex",
    reportNumber = "FERMILAB-PUB-15-262-ND",
    doi = "10.1103/PhysRevLett.116.151806",
    journal = "Phys. Rev. Lett.",
    volume = "116",
    number = "15",
    pages = "151806",
    year = "2016"
}

@article{LSND:2001aii,
    author = "Aguilar, A. and others",
    collaboration = "LSND",
    title = "{Evidence for neutrino oscillations from the observation of $\bar{\nu}_e$ appearance in a $\bar{\nu}_\mu$
 beam}",
    eprint = "hep-ex/0104049",
    archivePrefix = "arXiv",
    doi = "10.1103/PhysRevD.64.112007",
    journal = "Phys. Rev. D",
    volume = "64",
    pages = "112007",
    year = "2001"
}

@article{MiniBooNE:2020pnu,
    author = "Aguilar-Arevalo, A. A. and others",
    collaboration = "MiniBooNE",
    title = "{Updated MiniBooNE neutrino oscillation results with increased data and new background studies}",
    eprint = "2006.16883",
    archivePrefix = "arXiv",
    primaryClass = "hep-ex",
    reportNumber = "LA-UR-20-29235, LA-UR-20-24753, FERMILAB-PUB-20-288-AD-ND",
    doi = "10.1103/PhysRevD.103.052002",
    journal = "Phys. Rev. D",
    volume = "103",
    number = "5",
    pages = "052002",
    year = "2021"
}

@article{Huber:2011wv,
    author = "Huber, Patrick",
    title = "{On the determination of anti-neutrino spectra from nuclear reactors}",
    eprint = "1106.0687",
    archivePrefix = "arXiv",
    primaryClass = "hep-ph",
    doi = "10.1103/PhysRevC.85.029901",
    journal = "Phys. Rev. C",
    volume = "84",
    pages = "024617",
    year = "2011",
    note = "[Erratum: Phys.Rev.C 85, 029901 (2012)]"
}

@article{Barinov:2022wfh,
    author = "Barinov, V. V. and others",
    title = "{Search for electron-neutrino transitions to sterile states in the BEST experiment}",
    eprint = "2201.07364",
    archivePrefix = "arXiv",
    primaryClass = "nucl-ex",
    doi = "10.1103/PhysRevC.105.065502",
    journal = "Phys. Rev. C",
    volume = "105",
    number = "6",
    pages = "065502",
    year = "2022"
}

@article{SAGE:1998fvr,
    author = "Abdurashitov, J. N. and others",
    collaboration = "SAGE",
    title = "{Measurement of the response of the Russian-American gallium experiment to neutrinos from a Cr-51 source}",
    eprint = "hep-ph/9803418",
    archivePrefix = "arXiv",
    reportNumber = "SAGE-98-03",
    doi = "10.1103/PhysRevC.59.2246",
    journal = "Phys. Rev. C",
    volume = "59",
    pages = "2246--2263",
    year = "1999"
}

@article{GALLEX:1997lja,
    author = "Hampel, W. and others",
    collaboration = "GALLEX",
    title = "{Final results of the Cr-51 neutrino source experiments in GALLEX}",
    reportNumber = "DAPNIA-SPP-97-26, MPI-H-V41-1997, BNL-64864, ROM2F-97-45, TUM-SFB-375-221",
    doi = "10.1016/S0370-2693(97)01562-1",
    journal = "Phys. Lett. B",
    volume = "420",
    pages = "114--126",
    year = "1998"
}

@article{ALEPH:2005ab,
    author = "Schael, S. and others",
    collaboration = "ALEPH, DELPHI, L3, OPAL, SLD, LEP Electroweak Working Group, SLD Electroweak Group, SLD Heavy Flavour Group",
    title = "{Precision electroweak measurements on the $Z$ resonance}",
    eprint = "hep-ex/0509008",
    archivePrefix = "arXiv",
    reportNumber = "SLAC-R-774",
    doi = "10.1016/j.physrep.2005.12.006",
    journal = "Phys. Rept.",
    volume = "427",
    pages = "257--454",
    year = "2006"
}

@article{KARMEN:2002zcm,
    author = "Armbruster, B. and others",
    collaboration = "KARMEN",
    title = "{Upper limits for neutrino oscillations muon-anti-neutrino ---\ensuremath{>} electron-anti-neutrino from muon decay at rest}",
    eprint = "hep-ex/0203021",
    archivePrefix = "arXiv",
    doi = "10.1103/PhysRevD.65.112001",
    journal = "Phys. Rev. D",
    volume = "65",
    pages = "112001",
    year = "2002"
}

@article{MINOS:2020iqj,
    author = "Adamson, P. and others",
    collaboration = "MINOS+, Daya Bay",
    title = "{Improved Constraints on Sterile Neutrino Mixing from Disappearance Searches in the MINOS, MINOS+, Daya Bay, and Bugey-3 Experiments}",
    eprint = "2002.00301",
    archivePrefix = "arXiv",
    primaryClass = "hep-ex",
    reportNumber = "FERMILAB-PUB-20-054-ND",
    doi = "10.1103/PhysRevLett.125.071801",
    journal = "Phys. Rev. Lett.",
    volume = "125",
    number = "7",
    pages = "071801",
    year = "2020"
}

@article{Adamson:2015dkw,
    author = "Adamson, P. and others",
    title = "{The NuMI Neutrino Beam}",
    eprint = "1507.06690",
    archivePrefix = "arXiv",
    primaryClass = "physics.acc-ph",
    reportNumber = "FERMILAB-PUB-15-253-AD-FESS-ND",
    doi = "10.1016/j.nima.2015.08.063",
    journal = "Nucl. Instrum. Meth. A",
    volume = "806",
    pages = "279--306",
    year = "2016"
}

@article{Mufson:2015kga,
    author = "Mufson, S. and others",
    title = "{Liquid scintillator production for the NOvA experiment}",
    eprint = "1504.04035",
    archivePrefix = "arXiv",
    primaryClass = "physics.ins-det",
    reportNumber = "FERMILAB-PUB-15-048-ND-PPD",
    doi = "10.1016/j.nima.2015.07.026",
    journal = "Nucl. Instrum. Meth. A",
    volume = "799",
    pages = "1--9",
    year = "2015"
}

@article{Andreopoulos:2009rq,
    author = "Andreopoulos, C. and others",
    title = "{The GENIE Neutrino Monte Carlo Generator}",
    eprint = "0905.2517",
    archivePrefix = "arXiv",
    primaryClass = "hep-ph",
    reportNumber = "FERMILAB-PUB-09-418-CD",
    doi = "10.1016/j.nima.2009.12.009",
    journal = "Nucl. Instrum. Meth. A",
    volume = "614",
    pages = "87--104",
    year = "2010"
}

@article{Andreopoulos:2015wxa,
    author = "Andreopoulos, Costas and Barry, Christopher and Dytman, Steve and Gallagher, Hugh and Golan, Tomasz and Hatcher, Robert and Perdue, Gabriel and Yarba, Julia",
    title = "{The GENIE Neutrino Monte Carlo Generator: Physics and User Manual}",
    eprint = "1510.05494",
    archivePrefix = "arXiv",
    primaryClass = "hep-ph",
    reportNumber = "FERMILAB-FN-1004-CD",
    month = "10",
    year = "2015",
    journal = ""
}

@article{NOvA:2022wnj,
    author = "Acero, M. A. and others",
    collaboration = "NOvA",
    title = "{The Profiled Feldman-Cousins technique for confidence interval construction in the presence of nuisance parameters}",
    eprint = "2207.14353",
    archivePrefix = "arXiv",
    primaryClass = "hep-ex",
    reportNumber = "FERMILAB-PUB-22-476-ND",
    month = "7",
    year = "2022",
    journal=""
}

@article{Dydak:1983zq,
    author = "Dydak, F. and others",
    title = "{A Search for Muon-neutrino Oscillations in the Delta m**2 Range 0.3-eV**2 to 90-eV**2}",
    reportNumber = "CERN-EP/83-152",
    doi = "10.1016/0370-2693(84)90688-9",
    journal = "Phys. Lett. B",
    volume = "134",
    pages = "281",
    year = "1984"
}

@article{CCFRNuTeV:1998gjj,
    author = "Naples, D. and others",
    collaboration = "CCFR/NuTeV",
    title = "{A High statistics search for neutrino(e) (anti-neutrino(e)) ---\ensuremath{>} neutrino(tau) (anti-neutrino(tau)) oscillations}",
    eprint = "hep-ex/9809023",
    archivePrefix = "arXiv",
    reportNumber = "NUMI-412",
    doi = "10.1103/PhysRevD.59.031101",
    journal = "Phys. Rev. D",
    volume = "59",
    pages = "031101",
    year = "1999"
}

@article{FERMILABE531:1986uhg,
    author = "Ushida, N. and others",
    collaboration = "FERMILAB E531",
    title = "{Limits to Muon-neutrino, electron-neutrino ---\ensuremath{>} tau-neutrino Oscillations and Muon-neutrino, electron-neutrino ---\ensuremath{>} tau- Direct Coupling}",
    doi = "10.1103/PhysRevLett.57.2897",
    journal = "Phys. Rev. Lett.",
    volume = "57",
    pages = "2897--2900",
    year = "1986"
}

@article{CHORUS:2007wlo,
    author = "Eskut, E. and others",
    collaboration = "CHORUS",
    title = "{Final results on nu(mu) ---\ensuremath{>} nu(tau) oscillation from the CHORUS experiment}",
    eprint = "0710.3361",
    archivePrefix = "arXiv",
    primaryClass = "hep-ex",
    doi = "10.1016/j.nuclphysb.2007.10.023",
    journal = "Nucl. Phys. B",
    volume = "793",
    pages = "326--343",
    year = "2008"
}

@article{NOMAD:2001xxt,
    author = "Astier, P and others",
    collaboration = "NOMAD",
    title = "{Final NOMAD results on muon-neutrino ---\ensuremath{>} tau-neutrino and electron-neutrino ---\ensuremath{>} tau-neutrino oscillations including a new search for tau-neutrino appearance using hadronic tau decays}",
    eprint = "hep-ex/0106102",
    archivePrefix = "arXiv",
    reportNumber = "CERN-EP-2001-043",
    doi = "10.1016/S0550-3213(01)00339-X",
    journal = "Nucl. Phys. B",
    volume = "611",
    pages = "3--39",
    year = "2001"
}

@article{OPERA:2019kzo,
    author = "Agafonova, N. and others",
    collaboration = "OPERA",
    title = "{Final results on neutrino oscillation parameters from the OPERA experiment in the CNGS beam}",
    eprint = "1904.05686",
    archivePrefix = "arXiv",
    primaryClass = "hep-ex",
    doi = "10.1103/PhysRevD.100.051301",
    journal = "Phys. Rev. D",
    volume = "100",
    number = "5",
    pages = "051301",
    year = "2019"
}

@article{Aurisano:2016jvx,
    author = "Aurisano, A. and Radovic, A. and Rocco, D. and Himmel, A. and Messier, M. D. and Niner, E. and Pawloski, G. and Psihas, F. and Sousa, A. and Vahle, P.",
    title = "{A Convolutional Neural Network Neutrino Event Classifier}",
    eprint = "1604.01444",
    archivePrefix = "arXiv",
    primaryClass = "hep-ex",
    reportNumber = "FERMILAB-PUB-16-082-ND",
    doi = "10.1088/1748-0221/11/09/P09001",
    journal = "JINST",
    volume = "11",
    number = "09",
    pages = "P09001",
    year = "2016"
}

@article{Psihas:2019ksa,
    author = "Psihas, F. and Niner, E. and Groh, M. and Murphy, R. and Aurisano, A. and Himmel, A. and Lang, K. and Messier, M. D. and Radovic, A. and Sousa, A.",
    title = "{Context-Enriched Identification of Particles with a Convolutional Network for Neutrino Events}",
    eprint = "1906.00713",
    archivePrefix = "arXiv",
    primaryClass = "physics.ins-det",
    reportNumber = "FERMILAB-PUB-19-258-PPD",
    doi = "10.1103/PhysRevD.100.073005",
    journal = "Phys. Rev. D",
    volume = "100",
    number = "7",
    pages = "073005",
    year = "2019"
}

@article{NOvA:2021nfi,
    author = "Acero, M. A. and others",
    collaboration = "NOvA",
    title = "{Improved measurement of neutrino oscillation parameters by the NOvA experiment}",
    eprint = "2108.08219",
    archivePrefix = "arXiv",
    primaryClass = "hep-ex",
    reportNumber = "FERMILAB-PUB-21-373-ND",
    doi = "10.1103/PhysRevD.106.032004",
    journal = "Phys. Rev. D",
    volume = "106",
    number = "3",
    pages = "032004",
    year = "2022"
}

@article{NOvA:2021smv,
    author = "Acero, M. A. and others",
    collaboration = "NOvA",
    title = "{Search for Active-Sterile Antineutrino Mixing Using Neutral-Current Interactions with the NOvA Experiment}",
    eprint = "2106.04673",
    archivePrefix = "arXiv",
    primaryClass = "hep-ex",
    reportNumber = "FERMILAB-PUB-21-271-ND",
    doi = "10.1103/PhysRevLett.127.201801",
    journal = "Phys. Rev. Lett.",
    volume = "127",
    number = "20",
    pages = "201801",
    year = "2021"
}

@article{Esteban:2020cvm,
    author = "Esteban, Ivan and Gonzalez-Garcia, M. C. and Maltoni, Michele and Schwetz, Thomas and Zhou, Albert",
    title = "{The fate of hints: updated global analysis of three-flavor neutrino oscillations}",
    eprint = "2007.14792",
    archivePrefix = "arXiv",
    primaryClass = "hep-ph",
    reportNumber = "IFT-UAM/CSIC-112, YITP-SB-2020-21",
    doi = "10.1007/JHEP09(2020)178",
    journal = "JHEP",
    volume = "09",
    pages = "178",
    year = "2020"
}

@article{DayaBay:2022orm,
    author = "An, F. P. and others",
    collaboration = "Daya Bay",
    title = "{Precision Measurement of Reactor Antineutrino Oscillation at Kilometer-Scale Baselines by Daya Bay}",
    eprint = "2211.14988",
    archivePrefix = "arXiv",
    primaryClass = "hep-ex",
    doi = "10.1103/PhysRevLett.130.161802",
    journal = "Phys. Rev. Lett.",
    volume = "130",
    number = "16",
    pages = "161802",
    year = "2023"
}

@article{Parke:2015goa,
    author = "Parke, Stephen and Ross-Lonergan, Mark",
    title = "{Unitarity and the three flavor neutrino mixing matrix}",
    eprint = "1508.05095",
    archivePrefix = "arXiv",
    primaryClass = "hep-ph",
    reportNumber = "IPPP-15-53, DCPT-15-106, FERMILAB-PUB-15-363-T",
    doi = "10.1103/PhysRevD.93.113009",
    journal = "Phys. Rev. D",
    volume = "93",
    number = "11",
    pages = "113009",
    year = "2016"
}

@article{Nieves:2011pp,
    author = "Nieves, J. and Ruiz Simo, I. and Vicente Vacas, M. J.",
    title = "{Inclusive Charged--Current Neutrino--Nucleus Reactions}",
    eprint = "1102.2777",
    archivePrefix = "arXiv",
    primaryClass = "hep-ph",
    doi = "10.1103/PhysRevC.83.045501",
    journal = "Phys. Rev. C",
    volume = "83",
    pages = "045501",
    year = "2011"
}

@article{Megias:2016fjk,
    author = "Megias, G. D. and Amaro, J. E. and Barbaro, M. B. and Caballero, J. A. and Donnelly, T. W. and Ruiz Simo, I.",
    title = "{Charged-current neutrino-nucleus reactions within the superscaling meson-exchange current approach}",
    eprint = "1607.08565",
    archivePrefix = "arXiv",
    primaryClass = "nucl-th",
    doi = "10.1103/PhysRevD.94.093004",
    journal = "Phys. Rev. D",
    volume = "94",
    number = "9",
    pages = "093004",
    year = "2016"
}

@article{Katori:2013eoa,
    author = "Katori, Teppei",
    editor = "Da Motta, H. and Morfin, Jorge G. and Sakuda, M.",
    title = "{Meson Exchange Current (MEC) Models in Neutrino Interaction Generators}",
    eprint = "1304.6014",
    archivePrefix = "arXiv",
    primaryClass = "nucl-th",
    doi = "10.1063/1.4919465",
    journal = "AIP Conf. Proc.",
    volume = "1663",
    number = "1",
    pages = "030001",
    year = "2015"
}

@article{IceCube:2020phf,
    author = "Aartsen, M. G. and others",
    collaboration = "IceCube",
    title = "{eV-Scale Sterile Neutrino Search Using Eight Years of Atmospheric Muon Neutrino Data from the IceCube Neutrino Observatory}",
    eprint = "2005.12942",
    archivePrefix = "arXiv",
    primaryClass = "hep-ex",
    doi = "10.1103/PhysRevLett.125.141801",
    journal = "Phys. Rev. Lett.",
    volume = "125",
    number = "14",
    pages = "141801",
    year = "2020"
}

@article{Super-Kamiokande:2014ndf,
    author = "Abe, K. and others",
    collaboration = "Super-Kamiokande",
    title = "{Limits on sterile neutrino mixing using atmospheric neutrinos in Super-Kamiokande}",
    eprint = "1410.2008",
    archivePrefix = "arXiv",
    primaryClass = "hep-ex",
    doi = "10.1103/PhysRevD.91.052019",
    journal = "Phys. Rev. D",
    volume = "91",
    pages = "052019",
    year = "2015"
}

@article{Stockdale:1984cg,
    author = "Stockdale, I. E. and others",
    title = "{Limits on Muon Neutrino Oscillations in the Mass Range 55-eV**2 \ensuremath{<} Delta m**2 \ensuremath{<} 800-eV**2}",
    reportNumber = "UR-860, COO-3065-367",
    doi = "10.1103/PhysRevLett.52.1384",
    journal = "Phys. Rev. Lett.",
    volume = "52",
    pages = "1384",
    year = "1984"
}

@article{T2K:2019efw,
    author = "Abe, K. and others",
    collaboration = "T2K",
    title = "{Search for light sterile neutrinos with the T2K far detector Super-Kamiokande at a baseline of 295 km}",
    eprint = "1902.06529",
    archivePrefix = "arXiv",
    primaryClass = "hep-ex",
    doi = "10.1103/PhysRevD.99.071103",
    journal = "Phys. Rev. D",
    volume = "99",
    number = "7",
    pages = "071103",
    year = "2019"
}

@article{SciBooNE:2011qyf,
    author = "Mahn, K. B. M. and others",
    collaboration = "SciBooNE, MiniBooNE",
    title = "{Dual baseline search for muon neutrino disappearance at $0.5 {\rm eV}^2 < \Delta m^2 < 40 {\rm eV}^2$}",
    eprint = "1106.5685",
    archivePrefix = "arXiv",
    primaryClass = "hep-ex",
    reportNumber = "FERMILAB-PUB-11-304-E",
    doi = "10.1103/PhysRevD.85.032007",
    journal = "Phys. Rev. D",
    volume = "85",
    pages = "032007",
    year = "2012"
}

@article{MINOS:2016viw,
    author = "Adamson, P. and others",
    collaboration = "MINOS",
    title = "{Search for Sterile Neutrinos Mixing with Muon Neutrinos in MINOS}",
    eprint = "1607.01176",
    archivePrefix = "arXiv",
    primaryClass = "hep-ex",
    reportNumber = "FERMILAB-PUB-16-233-ND",
    doi = "10.1103/PhysRevLett.117.151803",
    journal = "Phys. Rev. Lett.",
    volume = "117",
    number = "15",
    pages = "151803",
    year = "2016"
}

@article{IceCube:2017ivd,
    author = "Aartsen, M. G. and others",
    collaboration = "IceCube",
    title = "{Search for sterile neutrino mixing using three years of IceCube DeepCore data}",
    eprint = "1702.05160",
    archivePrefix = "arXiv",
    primaryClass = "hep-ex",
    doi = "10.1103/PhysRevD.95.112002",
    journal = "Phys. Rev. D",
    volume = "95",
    number = "11",
    pages = "112002",
    year = "2017"
}

@article{Caldwell:1993kn,
    author = "Caldwell, David O. and Mohapatra, Rabindra N.",
    title = "{Neutrino mass explanations of solar and atmospheric neutrino deficits and hot dark matter}",
    reportNumber = "UCSB-HEP-93-03, UMD-PP-93-166",
    doi = "10.1103/PhysRevD.48.3259",
    journal = "Phys. Rev. D",
    volume = "48",
    pages = "3259--3263",
    year = "1993"
}

@article{Peltoniemi:1993ec,
    author = "Peltoniemi, J. T. and Valle, J. W. F.",
    title = "{Reconciling dark matter, solar and atmospheric neutrinos}",
    eprint = "hep-ph/9302316",
    archivePrefix = "arXiv",
    reportNumber = "FTUV-93-04, IFIC-93-04",
    doi = "10.1016/0550-3213(93)90174-N",
    journal = "Nucl. Phys. B",
    volume = "406",
    pages = "409--422",
    year = "1993"
}

@article{Bilenky:1998dt,
    author = "Bilenky, Samoil M. and Giunti, C. and Grimus, W.",
    title = "{Phenomenology of neutrino oscillations}",
    eprint = "hep-ph/9812360",
    archivePrefix = "arXiv",
    reportNumber = "UWTHPH-1998-61, DFTT-69-98, KIAS-P98045, SFB-375-310, TUM-HEP-340-98",
    doi = "10.1016/S0146-6410(99)00092-7",
    journal = "Prog. Part. Nucl. Phys.",
    volume = "43",
    pages = "1--86",
    year = "1999"
}

@article{Barger:2000ch,
    author = "Barger, Vernon D. and Kayser, Boris and Learned, J. and Weiler, Thomas J. and Whisnant, K.",
    title = "{Fate of the sterile neutrino}",
    eprint = "hep-ph/0008019",
    archivePrefix = "arXiv",
    reportNumber = "MADPH-00-1187, VAND-TH-00-7, AMES-HET-00-11",
    doi = "10.1016/S0370-2693(00)00950-3",
    journal = "Phys. Lett. B",
    volume = "489",
    pages = "345--352",
    year = "2000"
}

@article{Goldman:2000sc,
    author = "Goldman, J. Terrance and Stephenson, Jr., G. J. and McKellar, B. H. J.",
    title = "{Implications of quark lepton symmetry for neutrino masses and oscillations}",
    eprint = "nucl-th/0002053",
    archivePrefix = "arXiv",
    reportNumber = "LA-UR-00-908, UM-P-2000-007",
    doi = "10.1016/S0217-7323(00)00042-6",
    journal = "Mod. Phys. Lett. A",
    volume = "15",
    pages = "439--444",
    year = "2000"
}

@article{GEANT4:2002zbu,
    author = "Agostinelli, S. and others",
    collaboration = "GEANT4",
    title = "{GEANT4--a simulation toolkit}",
    reportNumber = "SLAC-PUB-9350, FERMILAB-PUB-03-339, CERN-IT-2002-003",
    doi = "10.1016/S0168-9002(03)01368-8",
    journal = "Nucl. Instrum. Meth. A",
    volume = "506",
    pages = "250--303",
    year = "2003"
}

@article{NOvA:2020dll,
    author = "Acero, M. A. and others",
    collaboration = "NOvA",
    title = "{Supernova neutrino detection in NOvA}",
    eprint = "2005.07155",
    archivePrefix = "arXiv",
    primaryClass = "physics.ins-det",
    reportNumber = "FERMILAB-PUB-20-201-E",
    doi = "10.1088/1475-7516/2020/10/014",
    journal = "JCAP",
    volume = "10",
    pages = "014",
    year = "2020"
}

@article{MicroBooNE:2021tya,
    author = "Abratenko, P. and others",
    collaboration = "MicroBooNE",
    title = "{Search for an Excess of Electron Neutrino Interactions in MicroBooNE Using Multiple Final-State Topologies}",
    eprint = "2110.14054",
    archivePrefix = "arXiv",
    primaryClass = "hep-ex",
    reportNumber = "FERMILAB-PUB-21-500-ND",
    doi = "10.1103/PhysRevLett.128.241801",
    journal = "Phys. Rev. Lett.",
    volume = "128",
    number = "24",
    pages = "241801",
    year = "2022"
}

@article{IceCube:2024dlz,
    author = "Abbasi, R. and others",
    collaboration = "IceCube",
    title = "{Search for a light sterile neutrino with 7.5~years of IceCube DeepCore data}",
    eprint = "2407.01314",
    archivePrefix = "arXiv",
    primaryClass = "hep-ex",
    doi = "10.1103/PhysRevD.110.072007",
    journal = "Phys. Rev. D",
    volume = "110",
    number = "7",
    pages = "072007",
    year = "2024"
}

@article{IceCubeCollaboration:2024nle,
    author = "Abbasi, R. and others",
    collaboration = "(IceCube Collaboration){\ensuremath{\parallel}}, IceCube",
    title = "{Search for an eV-Scale Sterile Neutrino Using Improved High-Energy {\ensuremath{\nu}}{\ensuremath{\mu}} Event Reconstruction in IceCube}",
    eprint = "2405.08070",
    archivePrefix = "arXiv",
    primaryClass = "hep-ex",
    doi = "10.1103/PhysRevLett.133.201804",
    journal = "Phys. Rev. Lett.",
    volume = "133",
    number = "20",
    pages = "201804",
    year = "2024"
}

@article{NOvA:2024imi,
    author = "Acero, M. A. and others",
    collaboration = "NOvA",
    title = "{Dual-Baseline Search for Active-to-Sterile Neutrino Oscillations in NOvA}",
    eprint = "2409.04553",
    archivePrefix = "arXiv",
    primaryClass = "hep-ex",
    reportNumber = "FERMILAB-PUB-24-0555-PPD",
    doi = "10.1103/PhysRevLett.134.081804",
    journal = "Phys. Rev. Lett.",
    volume = "134",
    number = "8",
    pages = "081804",
    year = "2025"
}

@phdthesis{Oh:2025cpz,
    author = "Oh, Haejun",
    title = "{NOvA Sterile Neutrino Search with Additional Systematic Constraints}",
    note = "https://inspirehep.net/literature/3005964, FERMILAB-THESIS-2025-27",
    school = "Cincinnati U., RWC",
    month = "7",
    year = "2025"
}

@phdthesis{Chaudhary:2026gtx,
    author = "Chaudhary, Shivam",
    title = "{First Sterile Neutrino Search at NOvA Experiment Using both Neutrino and Anti-Neutrino Data}",
    note = "https://inspirehep.net/literature/3187999, FERMILAB-THESIS-2026-17",
    school = "Indian Inst. of Info. Tech. Guwahati",
    month = "7",
    year = "2026"

}

@article{MINOS:2010fgd,
    author = "Adamson, P. and others",
    collaboration = "MINOS",
    title = "{Search for sterile neutrino mixing in the MINOS long baseline experiment}",
    eprint = "1001.0336",
    archivePrefix = "arXiv",
    primaryClass = "hep-ex",
    reportNumber = "FERMILAB-PUB-09-650-E",
    doi = "10.1103/PhysRevD.81.052004",
    journal = "Phys. Rev. D",
    volume = "81",
    pages = "052004",
    year = "2010"
}

\end{document}